\documentclass[twocolumn,showpacs]{revtex4-1}

\usepackage{amssymb}
\usepackage{amsmath}
\usepackage{graphicx}

\def\to{\rightarrow}

\def\ASAS{{\it Astron. and Astrophys.} }  
 \def\CQG{{\it Class. Quantum
		Gravity} }
    
  \def\JP{{\it
		J. Phys.} }

 \def\PL{{\it Phys. Lett.} } \def\PR{{\it
		Phys. Rev.} } \def\PRL{{\it Phys. Rev. Lett.} }    
\def\RMP{{\it Rev. Mod. Phys.} }

 \def\frac#1#2{{\textstyle{{#1}\over
			{#2}}}} 
\def\lsim{\mathrel{\rlap{\lower4pt\hbox{\hskip1pt$\sim$}}
		\raise1pt\hbox{$<$}}}
\def\gsim{\mathrel{\rlap{\lower4pt\hbox{\hskip1pt$\sim$}}
		\raise1pt\hbox{$>$}}} \def\sqr#1#2{{\vcenter{\vbox{\hrule
				height.#2pt \hbox{\vrule width.#2pt height#1pt \kern#1pt \vrule
					width.#2pt} \hrule height.#2pt}}}}

\def\beq{\begin{equation}} \def\eeq{\end{equation}}
\def\beqa{\begin{eqnarray}} \def\eeqa{\end{eqnarray}}

\usepackage{hyperref}

\begin{document}

\title{Baryogenesis in Nonminimally Coupled $f(R)$ Theories}

\author{M. P. L. P. Ramos}
\email[]{up201200278@fc.up.pt}
\author{J. P\'aramos}
\email[]{jorge.paramos@fc.up.pt}
\affiliation{Departamento de F\'\i sica e Astronomia and Centro de F\'isica do Porto, \\Faculdade de Ci\^encias da Universidade do Porto,\\Rua do Campo Alegre 687, 4169-007 Porto, Portugal}

\date{\today}

\begin{abstract}
We generalize the mechanism for gravitational baryogensis in the context of $f(R)$ theories of gravity, including a nonminimal coupling between curvature and matter. In these models, the baryon asymmetry is generated through an effective coupling between the Ricci scalar curvature and the net baryon current that dynamically breaks Charge Conjugation, Parity and Time Reversal ($CPT$) invariance. We study the combinations of characteristic mass scales and exponents for both non-trivial functions present in the modified action functional and establish the allowed region for these parameters: we find that very small deviations from General Relativity are consistent with the observed baryon asymmetry and lead to temperatures compatible with the subsequent formation of the primordial abundances of light elements. In particular, we show the viability of a power-law nonminimal coupling function $f_2(R) \sim R^n$ with $0 < n\lesssim 0.078$ and determine its characteristic curvature scale.

\end{abstract}


\maketitle

\section{\label{sec:level1}Introduction}

One of the major mysteries of our Universe today is the non-vanishing baryon asymmetry, which is inferred from the observed baryon-to-entropy ratio \cite{Planck}, 
\begin{equation}
\label{obs}
\eta_S^{\text{obs}} \equiv {n_b \over s} \lesssim 9 \times 10^{-11}~,
\end{equation}
where $n_b$ and $s$ are the baryon particle number and entropy densities, respectively. In the context of General Relativity (GR), entropy conservation implies that the number of baryons is proportional to this ratio, $N_b =a^3 n_b \sim \eta_S$, where $V=a^3$ is the comoving volume. It remains constant once the baryon-violating interactions are turned off and it is determined by precise measurements of the cosmic microwave background radiation (CMB) anisotropy spectrum and the predictions for the light element abundances produced during Big Bang nucleosynthesis (BBN) \cite{BBNreview}.

Starting from a neutral and symmetric universe, we would naturally expect the same amount of matter and antimatter. However, we do not see any bodies of antimatter up to the scale of clusters of galaxies. Within the Solar System, only measurements of cosmic rays yield a flux of antiprotons of about $ n_{\bar{p}} / n_p \sim 10^{-4}$. This means that $\eta_{S}$ can be regarded as the effective asymmetry parameter, that is, the ratio of the net baryon number, $n_B \equiv n_b - n_{\bar{b}}$, to the entropy density.

Hence, some mechanism must be in place in order to generate the overabundance of baryons compared to antibaryons. Avoiding the out-of-equilibrium scenario, which was one of the necessary conditions for a non-vanishing baryon asymmetry proposed by Sakharov \cite{Sakharov}, Cohen and Kaplan proposed a mechanism for generating $\eta_S$ while preserving thermal equilibrium, which is generally called spontaneous baryogenesis \cite{CK}.

This can happen if CPT is not a valid symmetry in the early Universe: although the Standard Model of particle physics is CPT-invariant by construction; at early times, these symmetries might not have been already established. Then, an expanding universe at finite temperature could in principle violate both Lorentz invariance --- if, for example, some vector field acquires a vacuum expectation value, as predicted by certain string theories \cite{Lehnert} --- and time reversal, before it reaches an isotropic and homogeneous ground state. This is accomplished by the introduction of an effective coupling between the net baryonic current $J^\mu$ and a scalar field $\phi$, of the form $J^{\mu} \nabla_\mu\phi $. Then, spontaneous baryogenesis can explain the thermal generation of $\eta_S$ via the classical motion of a scalar field, added \textit{ad hoc}, with specific initial conditions.

Following this work, Davoudiasl considered an identical coupling, but used the Ricci scalar curvature $R$ instead of a scalar field \cite{Davoudiasl}: this so-called gravitational baryogenesis mechanism adds the term below to the action functional,
\begin{equation}
\label{inter}
{1 \over M_*^2}\int d^4x \sqrt{-g} J^\mu\nabla_\mu R~,
\end{equation}
where $M_*$ is the cutoff scale of the effective theory. If non-vanishing, this term explicitly violates CPT because it causes an energy shift between particles and antiparticles. In thermal equilibrium, this can be interpreted as an effective chemical potential for baryons and antibaryons, $\mu_b\equiv \dot{R}/M_*^2=-\mu_{\bar{b}}$. Notice that this interaction eventually becomes vanishingly small for a matter or radiation dominated Universe, since the scale factor behaves as a power-law, $a(t) \sim t^{2/3(1+\omega)}$, thus implying that the scalar curvature drops as $t^{-2}$ and $\dot{R} \sim t^{-3}$.

Using the Fermi-Dirac thermal distribution in the limiting case where $T \gg m_B$, we obtain the usual net baryon number density $n_B = g_b \mu_B T^2 / 6$, where $g_b \sim O(1)$ are the intrinsic degrees of freedom of baryons. The entropy density, derived also from equilibrium thermodynamics, is given by 
\begin{equation}
\label{defentr}
s = {2 \pi^2 \over 45} g_{* S}(T) T^3~,
\end{equation}
where $g_{* S}(T)$, the effective number of relativistic degrees of freedom contributing to the entropy, is approximately constant equal to $g_* \approx 107$ for most of the history of the Universe (the difference between these values being significant only at low temperatures, as neutrinos decouple from the thermal bath). 

The baryon number to entropy ratio is then
\begin{equation}
\label{eta and R}
\eta_{S} = {n_B \over s} \approx -{15 g_b \over 4\pi^2 g_*} {\dot{R} \over M_*^2 T}, \quad \text{at $T=T_D$}~,
\end{equation}
where $T_D$ is the temperature at which the baryon-violating interactions decouple.

Note that, despite the ``gravitational'' attribute, gravity itself does not play any special role in this type of model for baryogenesis: it is only the background seen by the net baryon current, which exists, in the first place, due to the existence of some unspecified baryon ($B$)-violating force. The coupling (\ref{inter}) then enlarges the $B$-asymmetry until its present observational value, which is fixed when these forces decouple.

In GR, the Ricci scalar curvature is proportional to the trace of the energy-momentum tensor of matter, $R \sim T$. If the latter behaves as a perfect fluid with an equation of state (EOS) parameter $\omega=p/ \rho$ relating the pressure $p$ and energy density $\rho$, this reads $ R \sim (1-3\omega)\rho$. As such, in the radiation dominated epoch, characterized by $\omega = 1/3$, $R = \dot{R}$ vanishes and no net baryon number asymmetry can be generated. Modified gravity theories can easily avoid this, as the ensuing modified equations of motion can lead to very different relations between the scalar curvature and $T$ and evade this limitation.

In the past decades, among the different approaches to generalize GR, $f(R)$ theories have received a growing attention, being able to explain large scale structure and the current accelerating phase of the universe without the need for dark matter or dark energy (see Ref. \cite{Felice} for a thorough review). In this context, gravitational baryogenesis may occur, provided the form of the function $f(R)$ is nearly linear \cite{Lambiase}.

In an attempt to generalize $f(R)$ theories, one can include a nonminimal coupling (NMC) between matter and curvature \cite{early,NMC}: amongst other features, a NMC can also account for dark matter \cite{dm} and dark energy \cite{de} and give rise to the non-conservation of the energy-momentum tensor \cite{noncons} (see Ref. \cite{reviewNMC} for a review). 

The purpose of the present work is to investigate how an NMC can impact gravitational baryogenesis and determine how the correct value for the baryon asymmetry constraints its parameter space, thus extending the previous work reported in Ref. \cite{Lambiase}. It is divided as follows: in Section II, the model under scrutiny is presented. Section III then proceeds to detail the ensuing gravitational baryogenesis mechanism and extract the relevant quantities in terms of the model parameters. Section IV uses the experimental constraints on the latter to constraint the parameter space of the model, and conclusions are finally presented.

\section{\label{sec:levelII}The Model}

The action functional of a nonminimally coupled $f(R)$ theory reads
\begin{equation}
\label{action}
S= \int [\kappa f_1(R)+f_2(R)\mathcal{L}]\sqrt{-g}d^4x~,
\end{equation}
where $\kappa = (8\pi G)^{-1/2} = M_P/2$ and $M_P \approx 2.4\times 10^{18}$ GeV is the reduced Planck mass scale, $f_{i}(R)$ are arbitrary functions of the scalar curvature $R$, $g$ is the metric determinant and $\mathcal{L}$ the matter Lagrangian density; the standard Einstein-Hilbert action with a Cosmological Constant is obtained by taking $f_{1}(R) = R-2\Lambda$ and $f_{2}(R) = 1$, while  $f(R)$ theories are recovered by setting $f_1(R)= f(R)$ and $f_2(R) = 1$.

There is an equivalence between the above action and that of a two-scalar field model \cite{scalarNMC}, similarly to what occurs in $f(R)$ theories \cite{scalarfR}: one of these scalar fields is dynamically identified with the scalar curvature, so that the transition from spontaneous to gravitational baryogenesis appears quite naturally.

The field equations are obtained by imposing a null variation of the action with respect to the metric,
\begin{equation}
\label{feqs}
F R_{\mu\nu} = {1 \over 2}f_2T_{\mu\nu} + ( \nabla_{\mu}\nabla_{\nu} - g_{\mu\nu}\Box)F + {1\over 2} g_{\mu\nu} \kappa f_1~,
\end{equation}
where $F \equiv \kappa f'_{1}(R)+f'_{2}(R)\mathcal{L}(t)$ is defined for convenience, and the energy-momentum tensor is given by the usual expression,
\begin{equation}
T_{\mu\nu}= -{2 \over \sqrt{-g}}{\delta(\sqrt{-g}\mathcal{L}) \over \delta g^{\mu\nu}}~.
\end{equation}

The modified field Eqs. (\ref{feqs}) and the Bianchi identities imply that the energy-momentum tensor is no longer (covariantly) conserved:
\begin{equation}
\label{NC}
\nabla_{\mu}T^{\mu\nu} = {f'_2 \over f_2}(g^{\mu\nu}\mathcal{L} - T^{\mu\nu})\nabla_{\mu}R~.
\end{equation}
Following the equivalence with a two-scalar field model \cite{scalarNMC}, this may be recast as an energy exchange between matter and the former \cite{Harko}.

\subsection{Cosmology}

We consider a flat universe with a Friedmann-Robertson-Walker (FRW) metric
\begin{equation}
\label{FRW}
ds^{2} = dt^2 - a^2(t)dV^2~,
\end{equation}
where $a(t)$ is the scale factor and $dV$ is the volume element in comoving coordinates. Matter is assumed to behave as a perfect fluid, with energy-momentum tensor
\begin{equation}
T_{\mu\nu} = (p+\rho)u_{\mu}u_{\nu} -pg_{\mu\nu}~,
\end{equation}
where $u^{\mu}$ is its four-velocity.

The metric (\ref{FRW}) leads to the Ricci scalar curvature,
\begin{equation}
\label{curvature}
R = -6(\dot{H} + 2 H^2) ~,
\end{equation}
where $H(t) \equiv \dot{a}/a$ is the Hubble parameter.

The $tt$ component of Eqs. (\ref{feqs}) yields the modified Friedmann equation
\begin{equation}
\label{MFE}
-3(\dot{H} + H^2) F = {1\over 2}f_2\rho - 3H\dot{F} + {1 \over 2}\kappa f_1~,
\end{equation}

Likewise, the $rr$ component of the field equations reads
\begin{equation}
\label{RAY}
\left(\dot{H} + 3H^2\right)F = {1 \over 2}f_2p + \ddot{F} + 2 H\dot{F} - {1 \over 2}\kappa f_1~,
\end{equation}
while the trace of Eqs. (\ref{feqs}) yields
\begin{equation}
\label{TR}
3(\ddot{F} + 3H \dot{F}) - 6\left( \dot{H} + 2H^2 \right)F = {1 \over 2} f_2 (3p-\rho) + 2\kappa f_1~,
\end{equation}

Given that the NMC gives rise to an explicit dependence of the field equations on the Lagrangian density, we recall that radiation ({\it i.e.} matter characterized by an EOS parameter $\omega = 1/3$) is not only composed of photons, but also of relativistic particles. Moreover, due to the presence of the NMC, the on-shell degeneracy of the matter Lagrangian --- which is found in GR --- no longer holds. Following Ref. \cite{fluid}, we adopt the form $\mathcal{L}_B = \rho_B$ for the Lagrangian density of baryons, while for photons we have $\mathcal{L}_\gamma = -p_\gamma$ (notice the sign change due to the adopted metric signature).

For a perfect fluid, the energy and number densities and the pressure are related through $p = n (\partial \rho /\partial n) - \rho$, as obtained in Ref. \cite{Brown}. Using the EOS parameter, this condition leads to $\rho \sim n^{1+\omega} $, so that both relativistic baryons as well as photons scale as $\rho \sim n^{4/3}$, enabling us to write
\begin{eqnarray}
\rho & =& \rho_B + \rho_\gamma = \rho_\gamma\left(1+ {\rho_B \over  \rho_\gamma }\right) = \\ \nonumber 
& & \rho_ \gamma \left[1+\left({n_B \over n_\gamma }\right)^{4/3}\right] = \rho_\gamma\left(1+\eta^{4/3}\right)~,
\end{eqnarray}
where $\eta \equiv n_B / n_\gamma$ is the baryon to photon ratio. By the same token, the total Lagrangian is the sum of the Lagrangians of each species:
\begin{eqnarray}
\label{L}
\mathcal{L} & =& \rho_B -p_\gamma = \rho - \rho_\gamma - p_\gamma = \rho - {4 \over 3} \rho_\gamma \\
\nonumber & = & \rho\left[1-{4 \over 3}\left({1 \over 1+\eta^{4/3}}\right)\right]~.
\end{eqnarray}

Now that we have written the Lagrangian explicitly, Eq. (\ref{NC}) becomes:
\begin{eqnarray}
\label{derivative rho}
\dot{\rho} + 4{\dot{a} \over a}\rho & =&  {f'_2 \over f_2}\left[\rho_B - p_\gamma - (\rho_B + \rho_\gamma)\right]\dot{R} = \\ \nonumber
& &- {4 \over 3} {f_2' \over f_2} {\rho \over 1+\eta^{4/3}} \dot{R}~.
\end{eqnarray}
This can be directly integrated, considering $\eta = const.$, which is a good approximation if we neglect particle-antiparticle annihilation (below $T_{D}$) and ignore other processes, such as the production of photons in stars, as the majority is absorbed by nearby objects:
\begin{equation}
\label{integ}
\rho(t) = \rho_0 f_2(R(t))^{-{4 \over 3 ( 1+\eta^{4/3})}} a(t)^{-4} ~,
\end{equation}
where $\rho_0$ is the the energy density at an arbitrary time $t=t_0$. In the absence of a NMC, we recover the usual dependence $\rho \sim a^{-4}$ for a radiation dominated universe.

\section{Gravitational Baryogenesis}

To determine the cosmological dynamics depicted in the previous section, we now make the {\it Ansatz} that the scale factor evolves as a power-law, $a(t) \sim t^{\alpha}$ (with $\alpha >0$), so that 
\begin{equation}
\label{RicciScalar}
H(t) = {\alpha \over t} ~~~~,~~~~R(t) = 6{\alpha(1-2\alpha) \over t^2}~~.
\end{equation}

We also adopt power-law forms for the functions present in the action functional (\ref{action}),
\begin{eqnarray}\label{fchoice}
f_1(R) &=& R \left({|R| \over M_1^2}\right)^m~,
\\ f_2(R) &=& \left({|R| \over M_2^2}\right)^n~, \nonumber
\end{eqnarray}
where $M_i$ are characteristic mass scales; both $m$ and $n$ should be close to zero, in order to seek only slight deviations from GR.  Also, $1+m$ and $n$ must be greater than zero, so that no divergences in the action functional occur.

To account for other dynamics in the cosmological context, one usually assumes that both functions $f_{i}(R)$ can be written as a Laurent series,
\begin{equation}
f_{i}(R)=\sum_j \left({R \over R_{ij}}\right)^j~.	
\end{equation}
A single power-law model may be adopted in a particular context, such as dark energy or dark matter dominance, if the values of the scalar curvature relevant in each scenario imply that one of the terms of this Laurent series dominates the expansion, $f_{i}(R)\sim (R/R_{in})^{n}$, so that our treatment is compatible with the application of the model under consideration to other phenomena. In particular, in Subsection IV.A we discuss the competitive effect between $f_{1}(R)$, constrained to give the observable amount of matter asymmetry, and Starobinsky's model for inflation \cite{Starobinsky}.

We take the absolute value of the scalar curvature to allow it to be negative ({\it i.e.} $\alpha > 1/2$); alternatively, one could have considered negative values for $M_i^2$, although such notation is less appealing. Notice that, if we had adopted the metric signature $(-1,1,1,1)$, the curvature would change sign and we would be excluding the reciprocal region $0 < \alpha < 1/2$ --- effectively attributing physical significance to the metric signature. This detail was overlooked in Ref. \cite{Lambiase}, which as a result only studied half of the allowed parameter space.

Following Eq. {\ref{integ}), we have
\begin{equation}
\label{rhofr}
\rho(t) = \rho_0 \left({t \over t_0}\right)^{4\left({2 \over 3} {n \over 1+\eta^{4/3}} - \alpha \right)}~.
\end{equation}
Replacing Eqs. (\ref{RicciScalar}) and (\ref{rhofr}) into the modified field Eqs. (\ref{MFE}-\ref{TR}), we may obtain a relation between the exponents $\alpha$ and $n$. Considering Eq. (\ref{obs}), we neglect the extremely small value of the baryon-to-photon ratio, obtaining
\begin{equation}
\label{relation}
\alpha = {1 \over 2} \left( 1 + m + {n \over 3} {1 - 3 \eta^{4/3} \over 1 +  \eta^{4/3} } \right) \approx {1 \over 2} \left( 1 + m + {n \over 3} \right)~.
\end{equation}

Substituting into Eqs. (\ref{MFE}) or (\ref{RAY}) and solving for $\rho_{0}$, we obtain the energy density
\begin{equation}
\label{density mn}
\rho(t) = h_{mn}M_P^4 \left( {M_P\over M_1} \right)^{2m} \left( {M_2\over M_P} \right)^{2n} \left( M_P t\right)^{2(n-m-1)}~,
\end{equation}
defining the dimensionless quantity
\begin{eqnarray}
\nonumber h_{mn} &\equiv & { (3m + n)(3 + 3m + n) [3 + n - m (6 + 15 m + n)] \over 2[ 3 m (6 + 5 n) + n (3 + n)] } \\ &&  \times \left[ \left( 1 + m + {n \over 3} \right)\left| 3m + n \right| \right]^{m-n} ~.
\end{eqnarray}
\subsection{Baryon Asymmetry}

We now recall the usual result arising from statistical physics,
\begin{equation}
\label{deftemp}
\rho = {\pi^2 \over 30}g_* T^4~,
\end{equation}
and explicitly determine how temperature evolves,
\begin{eqnarray}
\label{temperature}
\nonumber T &=& M_P\left({30 \over g_* \pi^2} h_{mn}\right)^{1/4} \left( {M_P\over M_1} \right)^{m/2} \left( {M_2\over M_P} \right)^{n/2} \times \\ && \left( M_P t\right)^{(n-m-1)/2}~.
\end{eqnarray}

From definition (\ref{curvature}), it follows that
\begin{equation}
\dot{R} = {12(2\alpha-1)\alpha \over t^3}~.
\end{equation}
and the net baryon asymmetry can be written as
\begin{equation}
\label{eta parameter}
\eta_S \approx {g_b \over g_* } {45 \over \pi^2} {\alpha (2\alpha-1) \over t_D ^3 T_D M_*^2}~,
\end{equation} 
where $t_D$ is the decoupling time, at which the baryon violating interactions go out of equilibrium. Notice that the above can become negative, signalling the excess production of anti-matter: this could be corrected by changing the sign of the interaction term (\ref{inter}).

Inverting Eq. (\ref{temperature}), we obtain the relation $t=t(T)$, which we insert into Eq. (\ref{eta parameter}) so as to obtain
\begin{eqnarray}
\label{ass}
\eta_S &\lesssim & {5 \over 2\pi^2} { g_b \over g_* } l_{mn} \left({M_P \over M_*}\right)^2 \left({T_D \over M_P}\right) ^{5-m+n \over 1+m-n} \times \\ \nonumber && \left[ \pi \sqrt{ g_* \over 30 }  \left( {M_1\over M_P} \right)^m \left( {M_2\over M_P} \right)^{-n} \right]^{3/(1+m-n)}~,
\end{eqnarray}
where we have defined the dimensionless quantity
\begin{equation}
	l_{mn} \equiv ( 3 + 3m + n ) ( 3m + n ) (h_{mn})^{3/2(n-m-1)}~.
\end{equation}

As pointed out previously, it is natural to expect an operator such as (\ref{inter}) in the low effective field theory, if the cutoff scale $M_*$ is of the order the reduced Planck mass $M_P$. For this choice of $M_*$, the baryon asymmetry generated can be sufficiently large for $T_D = M_I$ \cite{Davoudiasl}, where $M_I \approx 2\times 10^{16}$ GeV is the upper bound on the energy scale of inflation, as placed by the Wilkinson Microwave Anisotropy Probe (WMAP) three-year data set \cite{inflationary bounds}. It is crucial that $T_D$ is placed after inflation, so that the asymmetry fixed once $B$-violating interactions decouple is not diluted by the ensuing exponential growth.

As hinted from the expression above, $\eta_{S}$ is very sensitive to the inflationary energy scale $M_I \sim T_D$: therefore, it is relevant to study numerically how the constraints on the exponents $(n,m)$ and mass scales $M_i$ are affected by the choice of the decoupling temperature $T_D$.

\subsection{Entropy Conservation}

As the baryon number to entropy ratio is paramount to our study, it is relevant to assess how the non-conservation law (\ref{NC}) for the energy-momentum tensor may affect adiabaticity \cite{Harko,matter creation}. In standard cosmology, the total entropy does not change as the Universe expands: since we know that, at low energies, there are no decays in which baryon number is created or destroyed, the baryon number to entropy ratio $\eta_S$ is constant. Similarly, once large scale annihilation processes have ended, the baryon to photon ratio $\eta$ is also constant, and both quantities can be swiftly related.

To assess the impact of a NMC, we resort to the first law of thermodynamics,
\begin{equation}
TdS = dE + pdV~,
\end{equation}
where $E = \rho (aL)^3$ and  $ S = s (aL)^3$ are the internal energy and entropy contained in an arbitrary comoving volume of size $L$, respectively, so that
\begin{eqnarray}
	T dS &=& d(\rho a^3) + p d(a^3) \to \\ \nonumber
	{T \over a^3} \dot{S} &=& \dot{\rho} + 4H \rho ~,
\end{eqnarray}

 Using  Eqs. (\ref{defentr}), (\ref{derivative rho}) and (\ref{deftemp}) leads to
\begin{equation}
(1 + \eta^{4/3}) {g_{*S}(T) \over g_*} {\dot{S} \over S} = - {f_2' \over f_2} \dot{R} ~.
\end{equation}
Assuming the baryon to photon ratio to be approximately constant, as discussed above, and taking $g_{*S}(T) \sim g_*$ allows us to directly integrate the above, obtaining
\begin{equation}
	 S(t) \sim f_2(R(t))^{ - {1 \over 1 + \eta^{4/3}}} \approx f_2(R(t))^{-1}~.
\end{equation}
so that the entropy remains constant in the absence of a NMC. Its variation can be neglected if it occurs at a rate much smaller than  the expansion rate of the Universe, 
\begin{equation}
\left| {\dot{S} \over S}\right| \approx \left| {f_2' \over f_2} \dot{R}\right| \ll H ~.
\end{equation}
Inserting Eqs. (\ref{relation}), (\ref{RicciScalar}) and (\ref{fchoice}) yields the simple condition $ | (11/3)n - m | \ll 1 $, which is naturally satisfied for the small perturbations $n, m \sim 0$ considered in the preceding section. Thus, we are led to conclude that the entropy remains approximately constant during  gravitational baryogenesis, so that $\eta \sim \eta_S \approx$ const.
\subsection{Big Bang Nucleosynthesis}

We now assess how the gravitational mechanism detailed in the previous sections can generate the correct amount of baryon asymmetry while maintaining compatibility with the typical temperatures $\sim 0.1-100$ MeV of BBN, the next major phase in the early Universe. Standard Cosmology sets the starting point of BBN very close to $T \approx 1$ MeV, when the weak interactions freeze-out \cite{Kolb}; in our framework, this can be extended to a higher value, due to the modified Hubble parameter that is used to define the Universe expansion rate; and to a lower one - until about $0.1$ MeV - which is characteristic of the temperature at which the mass fractions of the primordial elements get close to unity. This serves to justify the range of typical temperatures  that we allow, taking into consideration the possible conditions that can delay the production of the abundances, like properties of the elements themselves (such as their binding energy) or the usual ``bottlenecks'' --- the fact that the lack of light elements can prevent the production of heavier ones (since they participate in the formation reactions).

Baryogenesis must strictly occur before BBN, so that the initial conditions are in place to build up the observed abundances produced at early time, when the energy and number density were dominated by relativistic particles. At this stage of the evolution of the Universe, protons and neutrons are kept in thermal equilibrium by weak interactions, due to their rapid collisions:
\begin{equation}
\label{reactions}
\begin{split}
& n \longleftrightarrow p + e^- + \bar{\nu_e}~,\\
& \nu_e + n \longleftrightarrow p + e^-~,\\
& e^+ + n \longleftrightarrow p + \bar{\nu_e} ~.
\end{split}
\end{equation}

The weak interaction rate $\Lambda(T)$ is determined from the conversion rates of protons into neutrons. It corresponds to the sum of the decay rates of each reaction in Ref. (\ref{reactions}), plus the inverse ones. At sufficiently high temperatures, it is given by
\begin{equation}
\label{constants}
\Lambda(T) \approx {7\pi  \over 60}(1+3g_A^2)G_F^2T^5~,
\end{equation}
where $G_F \approx 1.166 \times 10^{-5}$ GeV$^{-2}$ is the Fermi coupling constant and $g_A\approx 1.27$ is the nucleon's axial-vector coupling constant \cite{Kolb,Bernstein}.

We are interested in the freeze-out temperature, $T_f$, at which the baryons decouple from leptons.  To compute it, one has to equate the rate of the weak interactions to the expansion rate of the universe, $\Lambda(T) \sim H$ since, from then on, the weak interaction rates are comparatively slower and the primordial abundances start being produced.

Using $g_*=g_*^{\text{BBN}}=10.75$ and Eqs. (\ref{RicciScalar}) and (\ref{temperature}), it follows that
\begin{eqnarray}
\label{Tf}
\nonumber T_f & = & \Bigg[{1 \over \pi} \sqrt{{30 h_{mn}\over g_*^{\text{BBN}} } }\left({7\pi (1+3g_A^2) \over 10 ( 3 + 3m + n ) } G_F^2 M_P^4 \right)^{1+m-n} \\ &&  \times\left( {M_P\over M_1} \right)^m \left( {M_2\over M_P} \right)^n \Bigg]^{1/(5n-5m-3)} M_P~.
\end{eqnarray}

\section{Parameter Constraints}

Given the results obtained above, we first impose the following set of requirements for the allowed values of the exponents $(n,m)$:
\begin{itemize}
\item The density, given by Eq. (\ref{density mn}), must be positive defined;
\item We consider only small deviations from GR, $m \sim 0$ and $n \gtrsim 0$;
\item An expanding Universe requires that $\alpha>0$, so that $m> - (1+n/3) $.
\end{itemize}
Using Eq. (\ref{density mn}), this yields,
\begin{itemize}
\item For $\alpha>1/2$, the condition 
\begin{equation}
-{(3+n)n \over 18 + 15 n} < m < {\sqrt{216 + 72 n + n^2}- n \over 30} -{1 \over 5} ~,
\end{equation}
\item For $\alpha<1/2$, $m$ can only take negative values,
\begin{equation} - {\sqrt{216 + 72 n + n^2} + n \over 30} - {1 \over 5} <m<-{n \over 3}~.
\end{equation}	
\end{itemize}

\noindent We now ascertain what are the allowed values for the exponents $(m,n)$ and mass scales $M_i$ compatible with the observed amount of asymmetry $\eta_{S}^{\text{obs}}$ and with a freeze-out temperature in the range $[0.1,100]$ MeV. To do so, we equal $\eta_{S}$ to its observational value and solve Eq. (\ref{ass}) for the combination $\left({M_P / M_1}\right)^m \left({M_2 / M_P}\right)^n $. We replace this  into Eq. (\ref{Tf}) to finally obtain
\begin{eqnarray}
\label{Tfnew}
\nonumber T_f = T_D & \Bigg[ & \sqrt{ g_*^{\text{BBN}} \over g_*^{\text{Bar}} }\bigg[ { 400 \over 343 \pi } { \eta_S^{\text{obs}} \over ( 1+3g_A^2 )^3 } {g_*^{\text{Bar}} \over g_b} { (3 + 3m + n)^2  \over 3m + n } \\  && \times\left({M_* \over G_F^3 T_D^7}\right)^2 \bigg]^{ 1+m-n \over 3} \Bigg]^{1/(3+5m-5n)}~.
\end{eqnarray}
\noindent where $g_*^{\text{Bar}} \approx 107$ corresponds to the relativistic degrees of freedom of species at $T=T_D \sim 10^{16}$ GeV, when the full set of Standard Model particles is effectively massless.

Imposing the requirements outlined at the beginning of this section and the constraint $10^{-4} <  T_f < 10^{-1}$ GeV yields the allowed combinations of exponents $(n,m)$ shown in Fig. \ref{allowedregion}, for different choices of the decoupling temperature $T_D$: in the past decade, the upper bounds on the inflation mass scale were refined from $3.3 \times 10^{16}$ GeV \cite{WMAP} to $2\times 10^{16}$ GeV \cite{inflationary bounds}.

As can be seen, the latter does not impact strongly on the overall shape of the allowed region --- indeed, a smaller $T_D$ only slightly shifts the allowed region into the lower right corner of the $(n,m)$ plane. Admitting $T_D = 2 \times 10^{16}$ GeV and considering a trivial NMC ($n=0$), we conclude that all values between $-0.07\lesssim m \lesssim 0.19$ are allowed, although a more precise measurement of $M_{I}$ could lower the upper limit of this range. Conversely, if we isolate the effect of the NMC (setting $m=0$), we find that any value of its exponent in the interval $0<n\lesssim 0.078$ is allowed.
\begin{figure}[ht]
\centering
\includegraphics[width=\columnwidth]{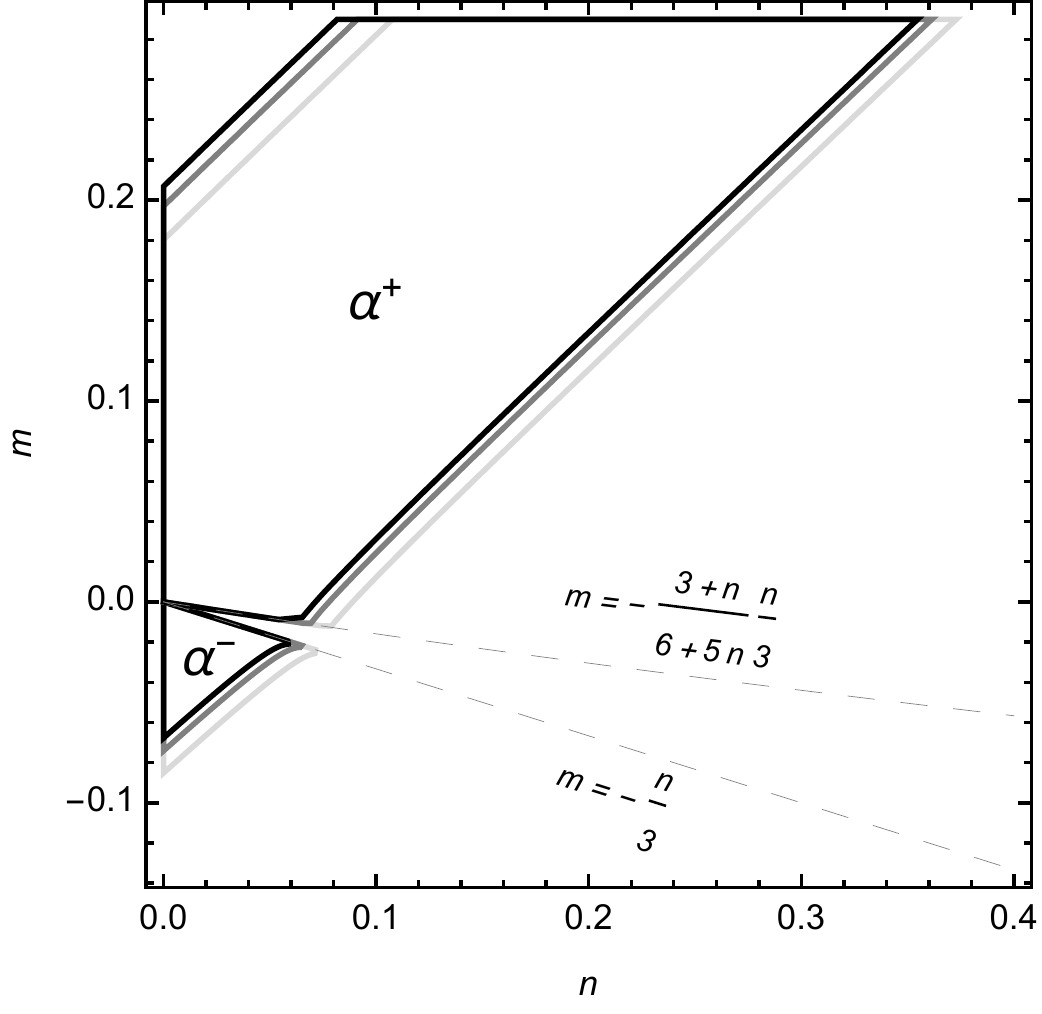}
\caption{Allowed regions for the exponents $(n,m)$: $\alpha^-$ and $\alpha^+$ correspond to $0<\alpha<1/2$ and $\alpha>1/2$, respectively. From  lighter to darker shade, $T_{D} = \{1, 2, 3\}\times 10^{16}$ GeV.}
\label{allowedregion}
\end{figure}
\begin{figure}[ht]
\centering
\includegraphics[width=\columnwidth]{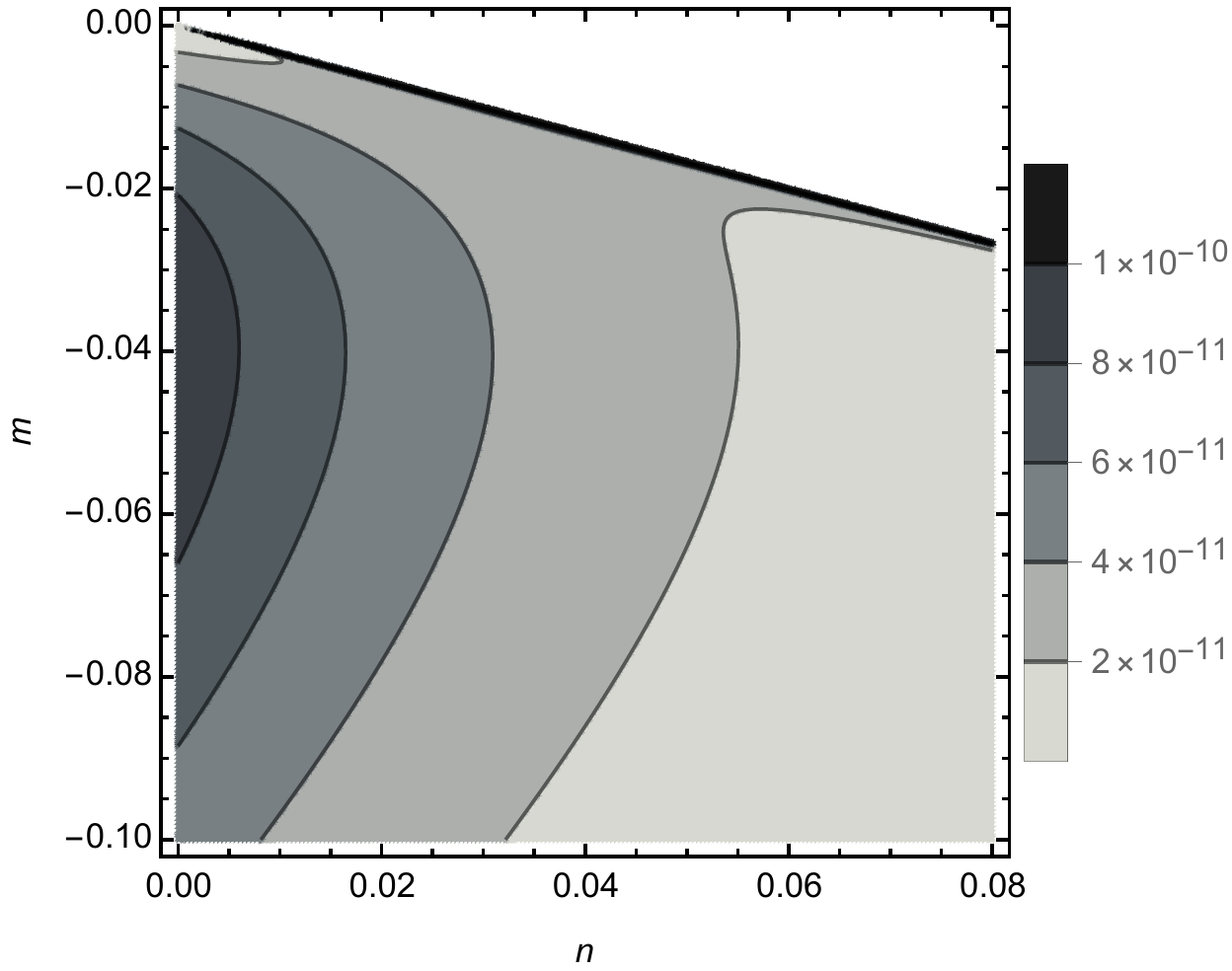}
\caption{Baryon-to-entropy ratio $\eta_S$ contour plot for the choice $M_1=M_2=M_P$ and $T_D = 3.3 \times 10^{16}$ GeV.}
\label{asymI3}
\end{figure}
\begin{figure}[ht]
\centering
\includegraphics[width=\columnwidth]{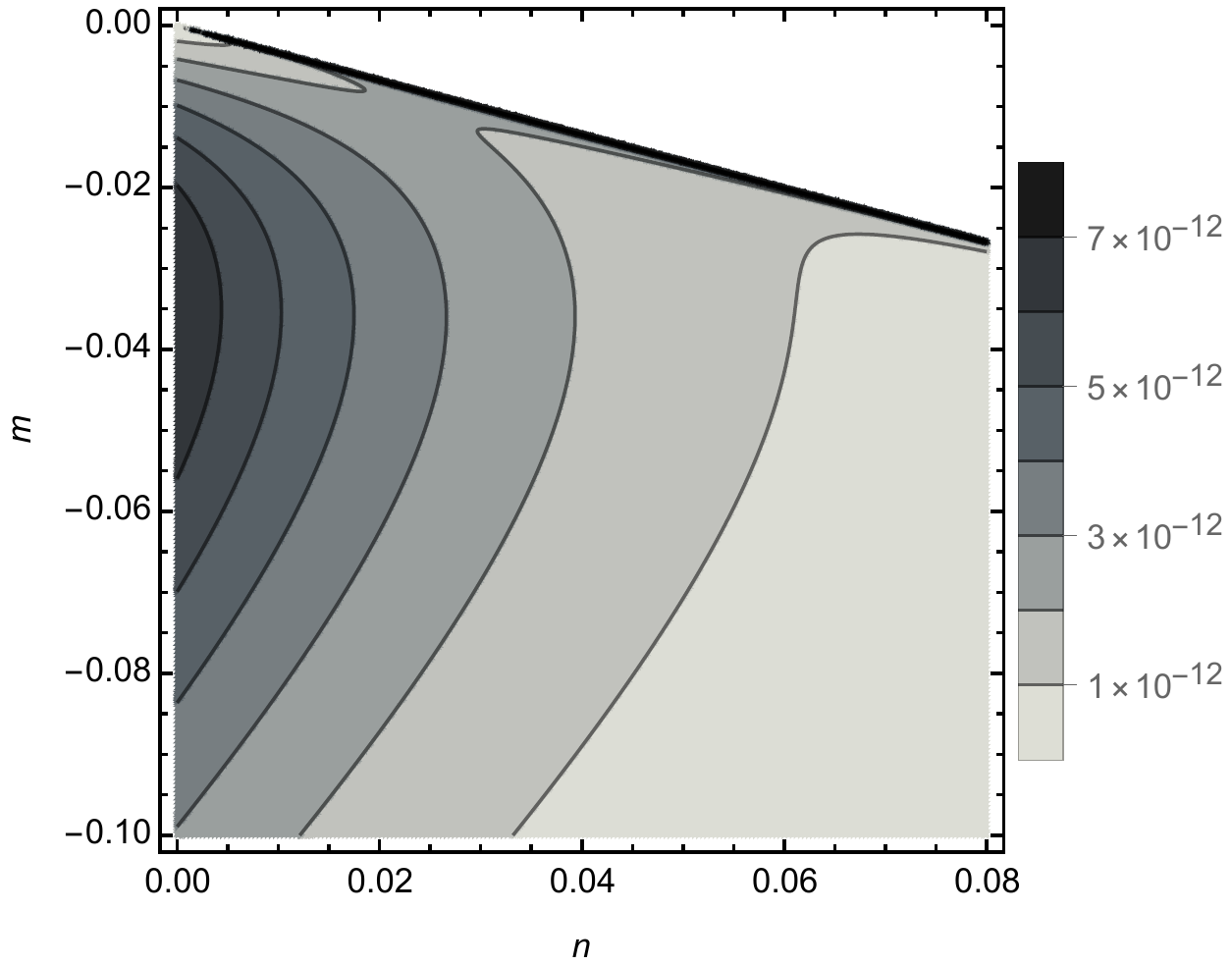}
\caption{Same as Fig. \ref{asymI3}, but with $T_D = 2 \times 10^{16}$ GeV.}
\label{asymI2}
\end{figure}

However, the decoupling temperature scale significantly alters the mass scales $M_i$ of our model: as an example, Fig. \ref{asymI3} presents the scenario when $M_1 = M_2 = M_P$, adopting the old bound for $M_I$ as the value of the decoupling temperature, $T_D = 3.3 \times 10^{16}$ GeV, showing that the right amount of baryon asymmetry $\eta_S \sim 10^{-10}$ can be attained. 

However, adopting the current bound for the inflationary energy scale, so that $T_D = 2 \times 10^{16}$ GeV, Fig. \ref{asymI2} shows that the ensuing baryon asymmetry is insufficient, $\eta_S \lesssim 7 \times 10^{-12}$, thus disallowing the possibility of both mass scales of our model lying at the Planck scale, $M_1=M_2=M_P$.

\subsection{Baryogenesis and $f(R)$ theories}
\label{baryof}

In this section we consider a minimal coupling $n=0$, aiming at generalizing the results obtained in Ref. \cite{Lambiase} for the exponent $m$ and the mass scale $M_1$. As can be checked from Fig. \ref{allowedregion}, we conclude that the former must lie within the range $ -0.07 \lesssim m \lesssim 0.19$, for $T_D = 2\times 10^{16}$ GeV.

We solve Eq. ({\ref{ass}}) for $M_1$ with $n=0$ to obtain it as a function of the exponent $m$, with a dependence on the inflationary and Planck mass scales of the form
\begin{equation}
\label{scaling}
	M_1^{3m} \sim M_P^3 M_*^{2(1+m)} T_D^{m-5} ~.
\end{equation}
Since $m \sim 0$, this mass scale turns out to be very sensitive to the inflationary mass scale $M_I \sim T_D$, as suggested in the preceding paragraph.

Fig. \ref{M1plot} shows this behaviour for different values of the decoupling temperature. For $T_D = 3.3 \times 10^{16}$ GeV, the maximum value $M_1 = 3.6 \times 10^{18}$ GeV $ = 1.5 M_P$ is attained for $m =-0.04$. 

Although other values for the exponent $m$ are permitted that correspond to sub-Planckian scales, $M_{1}\lesssim M_{P}$, we might ask what are the implications of having a characteristic mass scale $M_1 \sim M_P$ in other relevant cosmological scenarios: in particular, this could have some bearing on inflation, which occurs before the radiation dominance epoch is attained.

As such, we consider that inflation is driven by a quadratic curvature term, as posited by Starobinsky \cite{Starobinsky}, given by $f_{1}(R)=R+R^2/(6M_{S}^2)$, in which the linear term eventually causes inflation to end and $M_{S}\approx 10^{13}$ GeV \cite{Felice}: a De Sitter inflationary phase is attained as long as the quadratic term dominates the linear one, corresponding to the condition $R \gtrsim M_{S}^2$ is satisfied.

In this work, we thus adopt the form 
\begin{equation}
f_1(R) = R \left({|R| \over M_1^2}\right)^m + {R^2 \over 6M_S}~,	
\end{equation}
so that the quadratic term dominates the dynamics and enforces inflation if the curvature is high enough,
\begin{equation}
	R \gtrsim A_m M_{S}^{2} ~~~~,~~~~A_m \equiv \left[\sqrt{6}M_{S}/M_{1}(m)\right]^{2m/(1-m)}
\end{equation}
For all the values of the exponent $m$ considered in Fig. \ref{M1plot}, we find that $A_m \sim O(1)$, so that this lower bound is always of the same order of magnitude; as such, we conclude that the $R^{1+m}$ term responsible for baryogenesis has no impact on inflationary dynamics.

Drastically smaller mass scales are obtained for the current constraint $T_D = 2 \times 10^{16}$ GeV: $M_1$ is no longer of the order of $M_P$, but instead six orders of magnitude below, $M_1 \sim 10^{12}$ GeV.

This signals the strong dependence of the mass scale $M_1$ on the decoupling temperature $T_D$ for small values of the exponent $m$, as depicted in Fig. \ref{M1plot}: indeed, Eq. (\ref{scaling}) shows that for $m \sim 0$, $M_1 \sim T_D^{-5/3m}$; only for the unphysical case of very large deviations from GR is this alleviated, since $ |m| \gg 1$ implies that $M_1 \sim T_D^{1/3} $. 
\begin{figure}[ht]
\centering
\includegraphics[width=\columnwidth]{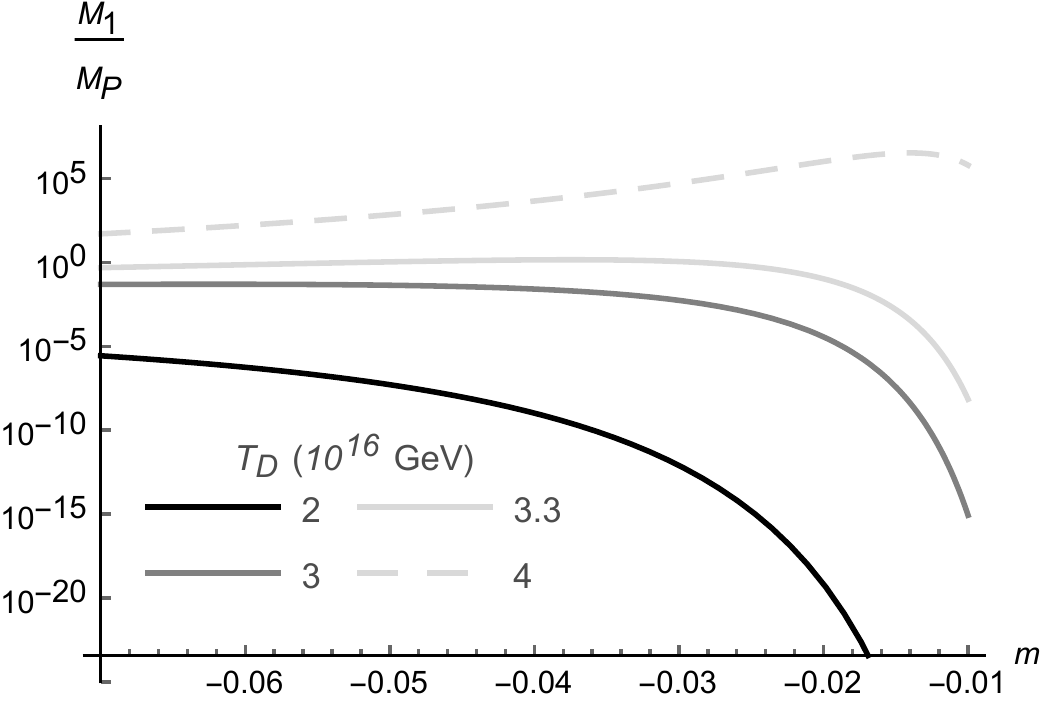}
\caption{Dependence of $M_1$ on the exponent $m$ for different choices of the decoupling temperature $T_D$.} 
\label{M1plot}
\end{figure}

\subsection{Gravitational Baryogenesis with a NMC}

A new result of our work is that a NMC can, by itself, induce gravitational baryogenesis. In order to isolate its effect, we set $m=0$ in the previous expressions, and conclude that the allowed values for $n$ compatible with BBN and the observed $B$ asymmetry lie within the interval $0 < n  \lesssim 0.078$.

Following the same argument as in the previous paragraph, we now consider both functions (\ref{ass}) and (\ref{Tf}) with $m=0$ in the regime where $\alpha>1/2$ and solve them for the characteristic mass scale $M_2$: this yields the scaling law $M_2^{3n} \sim M_P^{-3} M_*^{2(n-1)} T_D^{5+n}$ which, since $n \sim 0$, also leads to the conclusion that the mass scale of the NMC is very sensitive to the value adopted for the decoupling temperature, $ M_2 \sim T_D^{5 \over 3n}$. This is clearly shown in Fig. \ref{M2plot}: in particular, for $T_D = 3.3\times 10^{16}$ GeV we get the upper bound $n=0.070$ so that $ M_2 =6.5 \times 10^{13}$ GeV, while $T_D = 2 \times 10^{16}$ GeV gives a maximum value of $n= 0.078$ and a much smaller $M_2 = 2.5 \times 10^9 $ GeV.
\begin{figure}[ht]
\centering
\includegraphics[width=\columnwidth]{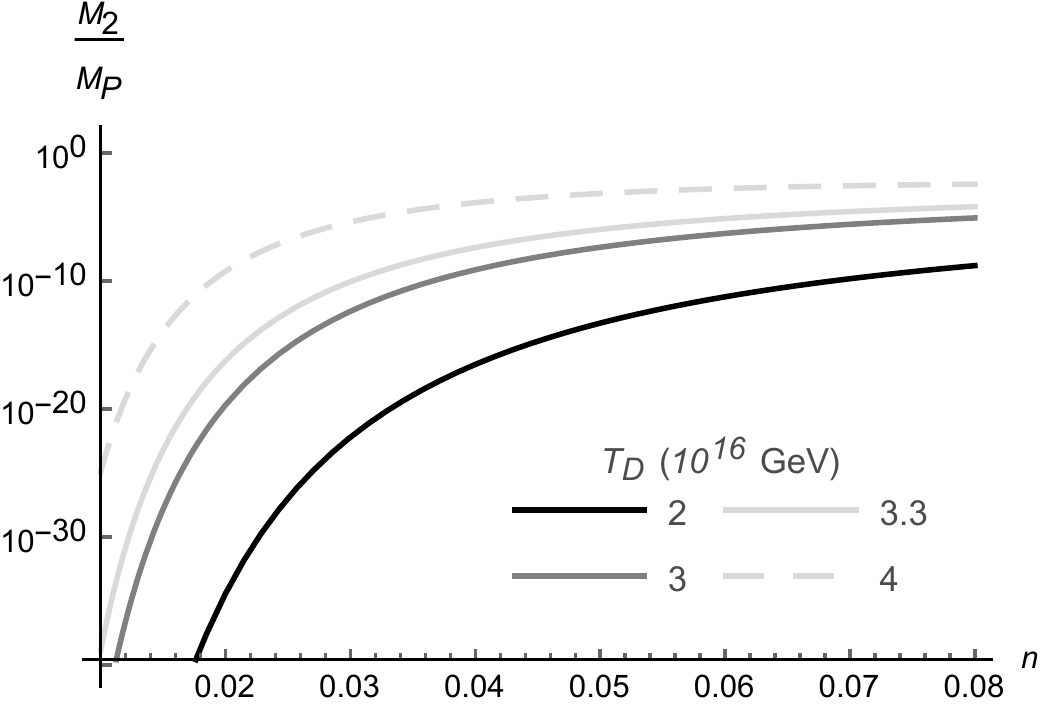}
\caption{Dependence of $M_2$ on the exponent $n$  for different choices of the decoupling temperature $T_D$.} 
\label{M2plot}
\end{figure}
\section{Discussion and Outlook}

\noindent In this work, we have studied how the inclusion of a NMC affects a mechanism for the generation of baryon asymmetry through an effective coupling between the net baryonic current and the derivative of the Ricci scalar, that dynamically breaks CPT invariance by inducing an energy shift between baryon and antibaryon thermal distributions.

Considering the non-conservation of the energy-momentum tensor, we integrated the first law of thermodynamics to read the evolution of the entropy: although it is not constant as in GR, we asserted that the time scale on which it varies significantly is much larger than the Hubble time as long as the constraint $|(11/3)n - m| \ll 1$ is kept: since small deviations from GR imply very small exponents $n$ and $m$, this is trivially fulfilled for all cases approached.

We have constrained the parameter space of a model with both a non-linear curvature term and a NMC: we showed that the observed amount of baryon asymmetry is attained with only a small deviation from GR, while keeping compatibility with the typical temperatures of Big Bang Nucleosynthesis. We also conclude that the characteristic mass scales are very sensitive to the  value for the decoupling temperature $T_D$ at which the baryon violation interactions go out of equilibrium and the baryon to entropy density becomes fixed --- which we admit to be of the same order of magnitude as the energy scale of inflation.

We have also extended the parameter space of Ref. \cite{Lambiase} by allowing for both positive and negative values of the Ricci scalar curvature, as its sign changes with the adopted metric signature and has no physical significance: we showed that a curvature term of the form $f_1(R) \sim R^{1+m}$ with $-0.07 \lesssim m \lesssim 0.19 $ can, by itself, generate the right amount of asymmetry. Although the allowed range of exponents is not very large (nor can it be, as we expect small deviations from GR, $m\sim 0$), the characteristic mass scale $M_1$ can vary significantly --- again depending crucially on the choice for $T_D$.

Finally, we find that a NMC of the form $f_{2}(R) \sim R^n$ is consistent with the observed $\eta$-parameter and BBN, with a small exponent in the range $0 < n \lesssim 0.078$. As expected from the application to $f(R)$ theories, the characteristic mass scale $M_2$ is again highly sensitive to the value of the decoupling temperature.

Future work could focus on the phenomenological consequences of the adopted form for the action functional and extract independent estimates on its characteristic mass scales $M_1$ and $M_2$: these could then be used to better restrict the allowed exponents $(n,m)$ and further assess what values of $T_D$ lead to the desired amount of baryogenesis --- and how these compare with the ever improving bounds on the energy scale of inflation.

\section*{Acknowledgements}\label{sec:acknowl}

J.P. acknowledges O. Bertolami and N. Mavromatos for fruitful discussions. The authors wish to thank the referees for her/his valuable comments and suggestions.


\begin{thebibliography}{00}
	
	\bibitem{Planck}P. A. R. Ade {\it et al.}, Planck Collaboration, \ASAS {\bf 594}, A13 (2016).
	
	\bibitem{BBNreview}B. Fields and S. Sarkar, \JP {\bf G 33}  1, (2006).
	
	\bibitem{Sakharov}A. D. Sakharov, {\it JETP Lett.} {\bf 5}, 24 (1967).
	
	\bibitem{CK}A. Cohen and D. Kaplan, \PL {\bf B 199}, 251 (1987).
	
	\bibitem{Lehnert}R. Lehnert, ``Handbook on Neutral Kaon Interferometry at a $\phi$-factory'', {\it Frascati Physics Series} {\bf 43} (2007).

	\bibitem{Davoudiasl}H. Davoudiasl, R. Kitano, G. D. Kribs, H. Murayama and P. J. Steinhardt, \PRL {\bf 93}, 201301 (2004)
	
	\bibitem{Felice}A. De Felice and S. Tsujikawa, {\it Liv. Rev. Rel.} {\bf 13}, 3 (2010).
	
	\bibitem{Lambiase}G. Lambiase and G. Scarpetta, \PR {\bf D 74}, 087504 (2006).
	
	\bibitem{early}L. Amendola and D. Tocchini-Valentini, \PR {\bf D 64}, 043509 (2001); S. Nojiri and S. D. Odintsov, {\it PoS WC} {\bf 2004}, 024 (2004); G. Allemandi, A. Borowiec, M. Francaviglia and S. D. Odintsov, \PR {\bf D 72}, 063505 (2005); T. Koivisto, \CQG {\bf 23}, 4289 (2006).

	\bibitem{NMC}O. Bertolami, C. G. B\"ohmer, T. Harko and F. S. N. Lobo, \PR {\bf D 75}, 104016 (2007).
			
	\bibitem{dm}O. Bertolami and J. P\'aramos, \textit{JCAP} {\bf 03}, 009 (2013); O. Bertolami, P. Fraz\~ao and J. P\'aramos, \PR {\bf D 86}, 044034 (2012).	

	\bibitem{de}O. Bertolami, P. Fraz\~ao and J. P\'aramos, \PR {\bf D 81}, 104046 (2010).
	
	\bibitem{noncons}T. P. Sotiriou and V. Faraoni, \CQG {\bf 25}, 205002 (2008).
		
	\bibitem{reviewNMC}O. Bertolami and J. P\'aramos, {\it Int. J. Geom. Meth. Mod. Phys.} 11, 1460003 (2014).

	\bibitem{scalarNMC}O. Bertolami and J. P\'aramos, \CQG {\bf 25}, 245017 (2008).

	\bibitem{scalarfR}P. Teyssandier and P. Tourranc, {\it J. Math. Phys.} {\bf 24}, 2793 (1983); H. Schmidt, \CQG {\bf 7}, 1023 (1990); D. Wands, \CQG {\bf 11}, 269 (1994).

	\bibitem{Harko}T. Harko, \PR {\bf D 90}, 044067 (2014).

	\bibitem{fluid}O. Bertolami, F. S. N. Lobo and J. P\'aramos, \PR {\bf D 78}, 064036 (2008).

	\bibitem{Brown}J. D. Brown,  \textit{Class. Quant. Grav.} {\bf 10} (1993)

	\bibitem{Starobinsky}A. Starobinsky, \PL {\bf B 91}, 99 (1980).
		
	\bibitem{inflationary bounds}W. H. Kinney, E. W. Kolb, A. Melchiorri and A. Riotto, \PR {\bf D 74}, 023502 (2006).

	\bibitem{matter creation}I. Prigogine, J. Geheniau, E. Gunzig and P. Nardone, \textit{Proc. Natl. Acad. Sci.} {\bf 85}, 7428 (1988).
	
	\bibitem{Kolb}E. W. Kolb, and M. S. Turner, ``The Early Universe'', Addison-Wesley Publishing Company (1989).
	
	\bibitem{Bernstein}J. Bernstein, L. S. Brown and G. Feinberg, \RMP {\bf  61}, 25 (1989)
	
	\bibitem{WMAP}H. V. Peiris et al., {\it Astrophys.J.Suppl.} {\bf 148}, 213 (2003).
	
\end{thebibliography}
\end{document}